\documentclass[final, 3p, times, authoryear]{elsarticle}

\usepackage{amssymb}

\usepackage{amsthm}

 \usepackage{lineno}

\usepackage{setspace}

\usepackage{float}

\biboptions{sort}

\usepackage{amsmath, amsthm, amssymb} 
\usepackage{bm} 

\usepackage{epsfig} 
\usepackage{graphicx} 

\def\bea{\begin{eqnarray}}
\def\eea{\end{eqnarray}}
\def\bean{\begin{eqnarray*}}
\def\eean{\end{eqnarray*}}
\renewcommand\eqref[1]{(\ref{#1})}

\def\R{\mathbb{R}}

\theoremstyle{plain}

\theoremstyle{definition}

\theoremstyle{remark}

\theoremstyle{note}

\usepackage{comment}
\usepackage{ulem} 
\usepackage[usenames,dvipsnames]{xcolor}
\definecolor{orange}{rgb}{1,0.5,0}
\definecolor{green}{rgb}{0.513,0.73,0.442}
\renewcommand\emph[1]{{\it #1}} 
\ifx \q \undefined
  \def \q[#1][#2][#3]{q_{#2}^{#3}\left(#1\right)}

\def\R{\mathbb{R}} 

\excludecomment{repeatIntro}
\excludecomment{styleRevisionFormat}
\excludecomment{styleRevisionColor}
\definecolor{NeonYellowWhiteOnBlack}{rgb}{0.016,0.009, 0.975}
\def\kwy#1{#1}
\begin{styleRevisionColor}
\def\kwy#1{\textcolor{NeonYellowWhiteOnBlack}{#1}}
\end{styleRevisionColor}

\definecolor{NeonBlueWhiteOnBlack}{rgb}{0.975, 0.016, 0.009}
\def\kwb#1{#1}
\begin{styleRevisionColor}
\def\kwb#1{\textcolor{NeonBlueWhiteOnBlack}{#1}}
\end{styleRevisionColor}

\definecolor{NeonOrangeWhiteOnBlack}{rgb}{0.0075, 0.44, 0.5525}
\def\kwo#1{#1}
\begin{styleRevisionColor}
\def\kwo#1{\textcolor{NeonOrangeWhiteOnBlack}{#1}}
\end{styleRevisionColor}

\def\kwdb#1{#1}
\begin{styleRevisionColor}
\def\kwdb#1{\textcolor{orange}{#1}}
\end{styleRevisionColor}

\def\kwp#1{#1}
\begin{styleRevisionColor}
\def\kwp#1{\textcolor{OliveGreen}{#1}}
\end{styleRevisionColor}

\definecolor{reproduction-gray}{gray}{0.65}

\def\paragraphTopicSentence#1{#1}
\begin{styleRevisionFormat}
\def\paragraphTopicSentence#1{
	\color{summarizing-gray}
	\emph{#1}
	\color{black}
}
\end{styleRevisionFormat}
\definecolor{summarizing-gray}{gray}{0.35}

\usepackage[geometry]{ifsym}
\newcommand{\Ring}[1]{\raisebox{-1pt}{\begin{tabular}{@{}c@{}}{\small #1}\\
      [-11.5pt]\BigCircle\end{tabular}}}
\newcounter{sectionSentenceCounter}[section] 
\newcounter{subsectionSentenceCounter}[subsection] 
\newcommand{\secsentence}{\refstepcounter{sectionSentenceCounter}\Ring\thesectionSentenceCounter}
\begin{document}

\begin{frontmatter}
%
%
%
%
\title{On idiosyncratic stochasticity of financial leverage effects}
%
\author{Carles Bret\'{o}} \fnref{label1}
\fntext[label1]{Tel. +34916245855; Fax:+34916249848}
\ead{carles.breto@uc3m.es}
\address{Departamento de Estad\'{i}stica and Instituto Flores de Lemus, Universidad Carlos III de Madrid, C/ Madrid 126, Getafe, 28903, Madrid, Spain}
%
%
\begin{abstract}
We model leverage as stochastic but independent of return shocks and of volatility and perform likelihood-based inference via the recently developed iterated filtering algorithm using S\&P500 data, contributing new evidence to the still slim empirical support for random leverage variation. 
%
%
%
%
%
\end{abstract}
%
%
\end{frontmatter}

%
%
\section{Introduction}
\label{sec:intro}
\def \SeemingTruth{\kwdb{Statistical modeling} of financial markets has become a key tool in the field of 
\kwp{finance}. In this 
\kwp{literature}, many statistical models focus on the dynamics of random 
\kwdb{volatilities of financial returns}. }
\def \SeemingTruthDetails{These 
\kwdb{volatilities} have been found to be correlated with shocks to returns and such correlations have been often referred to as  
\kwp{leverage effects}.  
\kwp{Leverage effects} have been considered by numerous studies, almost all of which assume that leverage is constant over time and which have provided ample 
\kwdb{evidence of its empirical relevance}. Part of this 
\kwdb{empirical evidence} has nevertheless suggested that leverage could vary randomly over time, as particularly supported by some recent data 
\kwp{analyses}. }
\def\ConditionOne{However, these 
\kwp{analyses} are very few so far, making empirical evidence regarding randomness (or lack thereof) of leverage still 
\kwdb{scant}. }
\def\ConsequenceOne{This 
\kwdb{scantness} is likely to discourage serious consideration of random leverage as an alternative to the widely accepted 
fixed leverage and failing to account for such randomness could distort inference (if leverage is indeed random) resulting, among other things, in 
\kwp{poorer hedging and risk management}. }
\def\SolutionOne{To help prevent such 
\kwp{poorer financial inferences} by casting light on the question of leverage randomness, this paper provides new 
\kwdb{empirical evidence} supporting random leverage. This 
\kwdb{evidence} is the central result of the paper and is obtained by considering a 
model in which leverage randomness is 
\kwp{idiosyncratic}, i.e., independent of shocks to returns and of volatility.  
Moreover, this empirical support for  
\kwp{idiosyncrasy} complements existing work on it that only addressed theoretical aspects.} 
%
%
%
%
%
%
%
%
\begin{styleRevisionFormat}
\def \seemingTruth{[Seeming truth:]}
\kwb{\seemingTruth}
\end{styleRevisionFormat}
\begin{styleRevisionFormat}
\secsentence\label{SeemingTruth}
\end{styleRevisionFormat}
\SeemingTruth
%
%
%
%
\begin{styleRevisionFormat}
\def \seemingTruthDetails{[Seeming truth details:]}
\kwb{\seemingTruthDetails}
\end{styleRevisionFormat}
\begin{styleRevisionFormat}
\secsentence\label{SeemingTruthDetails}
\end{styleRevisionFormat}
\SeemingTruthDetails
%
%
%
\begin{styleRevisionFormat}
\def \problem{[Problem condition:]}
\kwb{\problem}
\end{styleRevisionFormat}
\begin{styleRevisionFormat}
\secsentence\label{ConditionOne} 
\end{styleRevisionFormat}
\ConditionOne
%
%
%
%
\begin{styleRevisionFormat}
\def \soWhat{[So what?]}
\kwb{\soWhat}
\end{styleRevisionFormat}
\begin{styleRevisionFormat}
\secsentence\label{ConsequenceOne}
\end{styleRevisionFormat}
\ConsequenceOne
%
%
%
%
\begin{styleRevisionFormat}
\def \solution{[Solution:]}
\kwb{\solution}
\end{styleRevisionFormat}
\begin{styleRevisionFormat}
\secsentence\label{SolutionOne}
\end{styleRevisionFormat}
\SolutionOne
%
%
%
¥

%

\begin{repeatIntro}
\color{reproduction-gray} 
\begin{styleRevisionFormat}
\Ring{\ref{SeemingTruthDetails}} \seemingTruthDetails
\end{styleRevisionFormat}
\SeemingTruthDetails \\
\\ 
\color{black} 
\end{repeatIntro}
¥
\paragraphTopicSentence{\kwp{Leverage effects} have been historically modeled as \kwy{correlations} that are constant over time, although recent studies have \kwo{considered and supported the hypothesis of random variation} in this correlation, which is also the \kwdb{focus} of this paper.}
%
%
The study of \kwy{correlation} between volatility and shocks to returns is often traced back to \cite{black1976} and has gained popularity since then \citep[e.g.,][]{harvey1996, yu2005, francq2010, asai2011}. 
More recently, some studies have formally considered the possibility that this correlation be random and have found \kwo{evidence in favor of it}: \cite{bandi2012} considered both discrete- and continuous-time models but focused on jump-diffusion stochastic volatility models with leverage dynamics driven by spot volatility through a general function that was estimated non-parametrically relying on high-frequency S\&P500 data; 
and \cite{yu2012} proposed a discrete-time stochastic volatility model with white noise leverage driven by the sign and magnitude of shocks to returns and estimated it semi-parametrically based on multiple US return series (including S\&P500) relying on Markov chain Monte Carlo Bayesian methods. 
Also in this line of work, \cite{veraart2012} focused on the theoretical properties of continuous-time stochastic volatility models with leverage dynamics driven by parametric Jacobi processes independent of both shocks to returns and of volatility, but left for future work contributing further empirical evidence either supporting or failing to support both random leverage in general and idiosyncratic random leverage in particular, which is our \kwdb{main concern} in this paper. 
%
%
\begin{repeatIntro}
\\
\end{repeatIntro}

\begin{repeatIntro}
\color{reproduction-gray} 
\begin{styleRevisionFormat}
\Ring{\ref{ConditionOne}} \problem
\end{styleRevisionFormat}
\ConditionOne \\
\\ 
\color{black} 
\end{repeatIntro}
¥
\paragraphTopicSentence{The \kwdb{problem} that this paper aims to lighten is the limited current empirical evidence regarding randomness of leverage, which is aggravated both by a limited \kwy{theoretical exploration} of such randomness and by its complications in \kwo{inference} and which can \kwp{steer researchers away from considering} the recent support for random leverage.}
%
%
On one hand, much \kwy{theoretical work} characterizing fixed-leverage models exists and provides a haven for empirical analyses based on them (for a review of GARCH-type fixed-leverage models, see for example \citealp{francq2010}; for a review of stochastic volatility fixed-leverage models, see for example \citealp{asai2011}). 
On the other hand, theoretical work addressing random-leverage models seems to be rare and to consist mainly of the three references mentioned above (\citealp{bandi2012}; \citealp{veraart2012}; and \citealp{yu2012}). 
Random-leverage models may also imply a more complex \kwo{inference}, for example because of treating leverage as an unobserved variable \citep[in a state-space framework, as in][]{yu2012} or because of requiring high-frequency data (as in \citealp{bandi2012}; and \citealp{veraart2012}). 
These considerations about inference and theory are likely to complicate the flourishing of new studies addressing the empirical relevance of randomness in leverage, hence \kwp{detering its consideration as an alternative to the widely adopted fixed leverage}. 
%
%
\begin{repeatIntro}
\\
\end{repeatIntro}

\begin{repeatIntro}
\color{reproduction-gray} 
\begin{styleRevisionFormat}
\Ring{\ref{ConsequenceOne}} \soWhat
\end{styleRevisionFormat}
\ConsequenceOne \\
\\ 
\color{black} 
\end{repeatIntro}
¥
\paragraphTopicSentence{\kwp{Failure} to consider random leverage could induce both an \kwy{undue confidence in conclusions} and \kwo{biases} and would hence work against \kwb{prevention of poor} hedging and risk management, to which \kwdb{this paper} seeks to contribute.}
%
 %
\kwy{Misleading inference} can result from models that fix features that in fact vary. 
Examples of this include heteroskedastic disturbances in otherwise properly specified linear models \citep{white1980} and over-dispersed data in the context of generalized linear models \citep{mccullagh1989}, both of which result in downward biased standard errors (which lead to faulty inference). 
In general, similar \kwo{biases} are also likely to occur in any parameter of non-linear models \citep[as investigated for example in][]{he2010}, as a result of the model with the fixed parameter attempting to replicate the effects of the ignored parameter variation. 
Ignoring variation in leverage \kwb{could affect}, among others, hedging, due to the role of leverage in this operation \citep[as emphasized for example in][]{lai2011}, and risk management, through inaccurate inferences about (for example) value-at-risk. 
Contributing to prevent such poorer financial inferences is the \kwdb{ultimate goal of this paper}.
%
%
\begin{repeatIntro}
\\
\end{repeatIntro}

\begin{repeatIntro}
\color{reproduction-gray} 
\begin{styleRevisionFormat}
\Ring{\ref{SolutionOne}} \solution
\end{styleRevisionFormat}
\SolutionOne \\
\\ 
\color{black} 
\end{repeatIntro}
¥
\paragraphTopicSentence{The main \kwdb{contribution of this paper} is to provide new empirical evidence supporting random leverage based on a stochastic idiosyncratic \kwy{model} for it, in which \kwo{likelihood-based inference} is performed using \kwb{S\&P500 data}.}
%
%
The evidence comes from comparing two stochastic volatility \kwy{models}: the well-established, fixed-leverage, asymmetric autoregressive model of \cite{harvey1996} and one where the leverage parameter is assumed to follow a random walk (as defined in Section~\ref{sec:model}). 
Random walks are one of the simplest possible idiosyncratic dynamic models (since they only introduce one parameter: the noise variance), making them an interesting candidate for our goal, and can point to more complex models to be explored in future research. 
In addition, random walks have proved useful in similar contexts considered in the macroeconomics and financial time-varying parameter literature for several reasons \citep[see for example references in][]{muller2010}. 
These reasons include the ability to approximate breaks and that they can facilitate inference. 
\kwo{Inference} in this paper is implemented via the recently developed algorithm of iterated filtering for likelihood-based inference in general state-space models (as detailed in Section~\ref{sec:likelihood}) to analyze S\&P500 daily return data. 
When analyzing these \kwb{data} (in Section~\ref{sec:data}), the maximized likelihood for the fixed-leverage model is found to be lower than that for the random-walk model. 
This finding is consistent with interval estimates of the unobserved leverage process, which also suggest time variation. 
Both these results constitute the desired valuable evidence and (as discussed in Section~\ref{sec:discussion}) not only strengthen the still slim empirical support for random leverage (reported so far only for the non-idiosyncratic models of \citealp{yu2012} and of \citealp{bandi2012}) but also complement the theoretical considerations about idiosyncratic stochastic leverage of \cite{veraart2012}.  
%
%
%
\section{Stochastic volatility with random-walk leverage}
\label{sec:model}
Let $y_{t}$ be a financial rate of return in time period $t$ and $\sigma_t^2$ the volatility of $y_{t}$, both related by the discrete-time stochastic volatility model with the following state-space representation 
\begin{eqnarray}
\nonumber
y_t &=& \sigma_t^2 \epsilon_t = \exp{ \Big\{ h_t/2 \Big\} } \epsilon_t\\
\label{eqn:aarsv-long}
h_t &=& \mu_h(1-\phi) + \phi h_{t-1} + \eta_t \sigma_{\eta} \sqrt{1-\phi^2}\\
\nonumber
\text{Corr}\big[\eta_t, \epsilon_{t-1}\big] &=& \rho,
\end{eqnarray}
where $\{\epsilon_t\}$ and $\{\eta_t\}$ are Gaussian unit-variance white noise processes. 
In this model, both $y_{t}$ and $h_{t}$ are covariance stationary for $| \, \phi \, | < 1$. 
If we let $\eta_t =  \rho \epsilon_{t-1} + \omega_t \sqrt{1-\rho^2}$ with $\{\omega_t\}$ also being a Gaussian unit-variance white noise process that is independent of $\{\epsilon_t\}$, then $\{ \eta_t \}$ remains a Gaussian unit-variance white noise process with $\eta_t$ satisfying the required correlation with $\epsilon_{t-1}$. 
Hence, the set of equations~\eqref{eqn:aarsv-long} defines the same model as the set of equations 
\begin{eqnarray}
\nonumber
y_t &=& \sigma_t^2 \epsilon_t = \exp{ \Big\{ h_t/2 \Big\} } \epsilon_t\\
\label{eqn:aarsv}
h_t &=& \mu_h(1-\phi) + \phi h_{t-1} +  \sigma_{\eta} \sqrt{1-\phi^2} \Bigg( \rho \epsilon_{t-1} + \nu_t \sqrt{1-\rho^2} \Bigg)\\
\nonumber
&=& \mu_h(1-\phi) + \phi h_{t-1} + \beta \rho \exp{\Big\{-h_{t-1}/2 \Big\}} + \sigma_{\omega} \omega_t,
\end{eqnarray}
with $\beta= y_{t-1} \sigma_{\eta} \sqrt{1-\phi^2}$ and $\sigma_{\omega} = \sigma_{\eta} \sqrt{1-\phi^2} \sqrt{1-\rho^2}$. 
That sets of equations~\eqref{eqn:aarsv-long} and~\eqref{eqn:aarsv} are equivalent has often been pointed out \citep[see for example,][]{harvey1996, malik2011, yu2012}. 
The stochastic volatility model defined by set of equations~\eqref{eqn:aarsv} is a re-parameterization of the asymmetric stochastic volatility model of \cite{harvey1996} (so that neither the mean nor the variance of $h_{t}$ depend on $\phi$) and this will be the model that we will allude to when in the rest of the paper we refer to the fixed-leverage model. 
%

%
As an alternative to the fixed-leverage model, we consider letting an appropriate transformation of leverage follow a random walk. 
If such transformation is the Fisher transformation, leverage is allowed to vary over its natural range $[-1,1]$, whereas a logit transformation might be preferred if negativity is to be imposed (which we avoid given the existing evidence pointing to the possibility of positive correlations, as in \citealp{yu2012}). 
The Fisher-transformed leverage $\{ \rho_{t} \}$ process driven by the random-walk leverage factor process $\{ f_{t} \}$ is defined by equations
\begin{eqnarray}
\label{eqn:fisher}
\rho_t &=& \frac{e^{2 f_t} - 1}{e^{2 f_t} + 1}\\ 
\nonumber
f_t &=& f_{t-1} + \sigma_{\nu} \nu_t,
\end{eqnarray}
where $\{\nu_{t}\}$ is a Gaussian white noise process again independent of $\{\epsilon_{t}\}$ and of $\{\eta_{t}\}$. 
Replacing $\rho$ by~\eqref{eqn:fisher} in equations~\eqref{eqn:aarsv} gives the following state-space model equations
\begin{eqnarray}
\nonumber
y_t &=& \sigma_t^2 \epsilon_t = \exp{ \Big\{ h_t/2 \Big\} } \epsilon_t\\
\label{eqn:rw-aarsv}
h_t &=& \mu_h(1-\phi) + \phi h_{t-1} + \beta \frac{e^{2 (f_{t-1} + \sigma_{\nu} \nu_t)} - 1}{e^{2 (f_{t-1} + \sigma_{\nu} \nu_t)} + 1} \exp{\Big\{-h_{t-1}/2 \Big\}} + \sigma_{\omega} \omega_t \\
\nonumber
f_t &=& f_{t-1} + \sigma_{\nu} \nu_t, 
\end{eqnarray}
which define a model that we refer to as the random-walk leverage model. 
In spite of the random walk, both $E[\rho_{t}]$ and $V[\rho_{t}]$ are bounded, even if $\{\rho_{t}\}$ is not covariance stationary. 
The covariance stationarity of both $y_{t}$ and $h_{t}$ for $| \, \phi \, | < 1$ of the fixed-leverage model is retained by the random-walk leverage model, both of which belong to the family of non-linear, non-Gaussian state-space models. 
%
%
%

%
%
\section{Likelihood-based inference}
\label{sec:likelihood}
Non-linear, non-Gaussian state-space models have been studied extensively \citep[see for example][]{durbin2001}. 
In these models, it is not straighforward to check that the appealing, usual properties of maximum likelihood estimates hold, in part due to the likelihood being in general analytically intractable. 
These difficulties did not prevent early developments of simulation-based methods to perform likelihood-based inference in such models \citep{shephard1997, durbin1997}, which is on its own a challenge that has been taken on more recently \citep{liesenfeld2003, richard2007, jungbacker2007}. 
Another recent contribution, which we favor in this paper, is the iterated filtering algorithm. 
%

%
Iterated filtering has been successfully implemented to do inference in highly non-Gaussian, non-linear state-space models, mostly in a context of biological applications \citep{bhadra2011, breto2009, he2010, king2008, laneri2010, shrestha2011} but also in the continuous-time L\'{e}vy-driven stochastic volatility model of \cite{barndorff2001} \citep[see the discussion on][]{andrieu2010}.
The iterated filtering algorithm is based on a sequence of filtering operations which have been shown to converge to the maximum likelihood estimate \citep{ionides2006-pnas, ionides2011-aos}. 
The algorithm boils down to extending the model of interest by letting the model parameters that are to be estimated become artificially random. 
Then, starting at an initial guess for the estimate $\hat{\theta}_{0}$ and successively diminishing the artificial random perturbations to the parameters, the algorithm generates a sequence of updated parameter estimates $\hat{\theta}_{1}, \hat{\theta}_{2}, \ldots$ that converges to the maximum likelihood estimate. 
Iterated filtering has been implemented in this paper using the software package POMP \citep[Partially Observed Markov Processes,][]{king2010} written for the R statistical computing environment \citep{r10}. 
Iterated filtering can be implemented as long as the state-space model satisfies the usual conditions of state Markovianity and of measurement independence conditional on current states (which are met by our models, in spite of $y_{t-1}$ entering the equations for $h_{t}$ through $\beta$). 
Nevertheless, these two conditions are necessary but not sufficient to converge to a maximum. 
%

%
%

%
Convergence of iterated filtering to the maximum likelihood estimate has been shown under some regularity conditions on the likelihood surface and on the algorithmic parameters \citep[see][for details]{ionides2006-pnas, ionides2011-aos}. 
The algorithmic parameters include the number of iterations, the initial levels of artificial randomness and the speed at which this randomness diminishes from one iteration to the next (and also the number of particles if iterated filtering is implemented via particle filters, as in this paper).
The conditions on the likelihood surface are rather technical so, in practice, convergence is assessed via convergence diagnostic plots (included in~\ref{app:diagnostics}). 
These plots suggest an appropriate choice of algorithmic parameters and represent evidence that the likelihood has in fact been maximized in our analysis of S\&P500 data. 
%

%

%
%
%
\section{Empirical results}
\label{sec:data}
A sample of S\&P500 daily \kwp{returns} is analyzed in Table~1 and in Figure~1, which report \kwy{estimates and results} that broadly agree with the \kwo{related results} provided by \cite{bandi2012} and by \cite{yu2012} and that seem to \kwdb{support random and idiosyncratic leverage variation}. 
The \kwp{returns} were demeaned prior to the analysis and correspond to the 25-year period from January 4, 1988 to December 31, 2012. 
In this period, there are 6302 observations, which is comparable to the sample size used in \cite{bandi2012}, almost six times that used in \cite{yu2012}, and crucial to obtain the high estimate precision reported in Table~1. 
All the \kwy{estimates} in Table~1 seem to be statistically significant and they correspond to either parameters that are common to both models or specific to each model. 
On one hand, the estimates for common parameters $\mu_{h}, \phi$ and $\sigma_{\eta}$ are almost identical for both models (and do not appear to be statistically different). 
They are also in the usual range reported in the literature (after accounting for the fact that some analysis may not use our parameterization to remove $\phi$ from the marginal mean and variance of $h_{t}$, e.g., \citealp{yu2012}). 
On the other hand, the estimated correlation $\rho$ specific to the fixed-leverage model lies in the interval $(-0.7399, -0.5759)$. 
This interval is in the upper end of correlation strength usually reported but seems to \kwo{agree with the results} in \cite{yu2012} and in \cite{bandi2012}. 
On one hand, \cite{bandi2012} also find evidence of leverage being ``more negative than those generally found in the literature'' \citep[][page105, in light of their Fig.~4]{bandi2012} based on results for the 1990-2009 period (which is a large subset of the period we consider). 
On the other hand, \cite{yu2012} does not find such a strong correlation for the period 1985-1989 (which is a much shorter period ending one year after ours begins) and in which the interval estimate for $\rho$ is approximately $(-0.5431, -0.1951)$. 
This interval is nevertheless very similar to our filtering interval for $\rho_{t}$ of about $(-0.55,-0.10)$ reported around the period 1988-1999 in Figure~1. 
Figure~1 also supports a relatively stable leverage around $-0.4$ for the decade 1988-1997 (roughly, the first 2500 observations) followed by a progressive drop to about $-0.8$ for the last decade 2003-2012 (roughly, the last 2500 observations). 
This strong correlation in the last decade agrees with the findings reported in \cite{bandi2012}, as explained in the lines above. 
The smooth progression from around $-0.4$ to around $-0.8$ and the stability around these values for about a decade each are also consistent with the fixed-leverage correlation being estimated around $-0.6$. 
A final finding consistent with previous analysis that we point out is positive leverage. 
Positive leverage was found in \cite{yu2012} and, although very timidly, it also appears in Figure~1 within the filtering intervals of $\rho_{t}$ for only a few months in one of the early years in our 25-year sample. 
Providing a sound, deeper financial interpretation of these empirical findings is beyond the goal of this paper of \kwdb{throwing light on the issue of leverage randomness}. 
\begin{table}[ht]
\label{tab:results}
{\bf \caption{ \textnormal{Parameter and log-likelihood ($\ell$) estimates and standard errors (in parenthesis) for the fixed- and random-walk leverage models and S\&P500 data. 
These data were analyzed with the following algorithmic parameters:  150 iterations, an exponential decay of perturbations of $\alpha=0.978$ and 8,000 particles. 
70,000 particles were used to obtain the estimates of $\ell$, which result from taking the average of two likelihood evaluations from which we calculate the Monte Carlo standard error. 
Parameter standard errors were derived via a numerical approximation to the Hessian \citep[see][]{ionides2006-pnas}. 
They are computationally affordable and give a reasonable idea of the scale of uncertainty and can hence be used to construct approximate confidence intervals. 
More rigorous confidence intervals can be obtained via profile likelihoods. 
}}}
\centering
\begin{center}
\begin{tabular}{l | ccc | cc | c}
\hline \hline
Model & $\mu_{h}$ & $\phi$ & $\sigma_{\eta}$ & $\rho$ & $\sigma_{\nu}$  & $\ell$\\
\hline \hline
{Fixed leverage}  & -0.2506&  0.9805&  0.9003& -0.6579& --&-8416.44 \\    
& (0.0710)&  (0.0017)& (0.0375)& (0.0599)& --&(0.0410) \\
\hline
{Random-walk leverage}  & -0.2610 &      0.9818&    0.9222&  --& 0.0086& -8409.06 \\
&(0.0776)&    (0.0015)&  (0.0406)& --& (0.0013) &(0.1333) \\
\hline\hline
\end{tabular}
\end{center}
\end{table}

\begin{figure}[ht]
\label{fig:data}
\centering
\begin{minipage}[b]{0.49\linewidth}
  \includegraphics[width=1\linewidth]{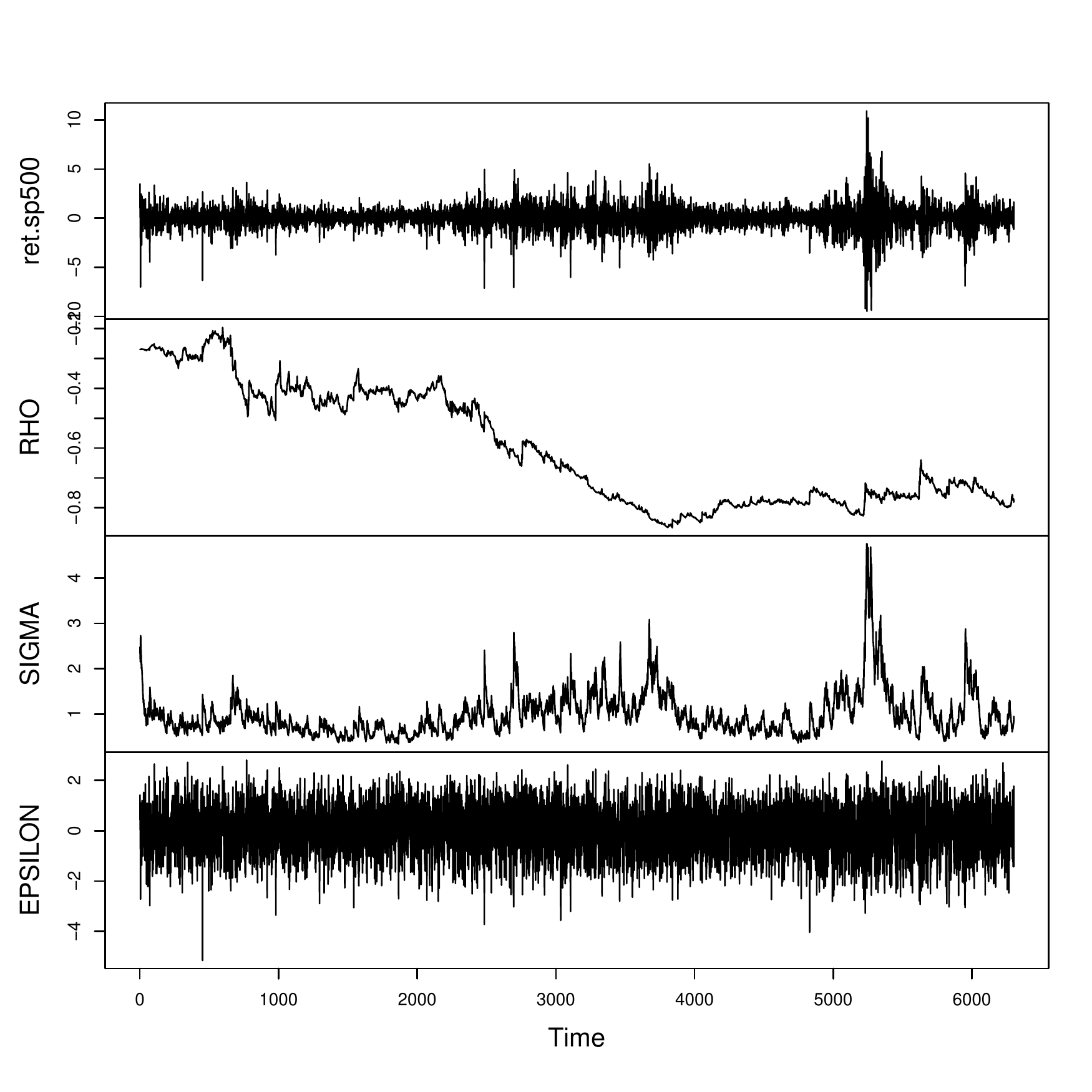}
\end{minipage}
\hspace{-0.3cm}
\begin{minipage}[b]{0.49\linewidth}
  \includegraphics[width=1\linewidth]{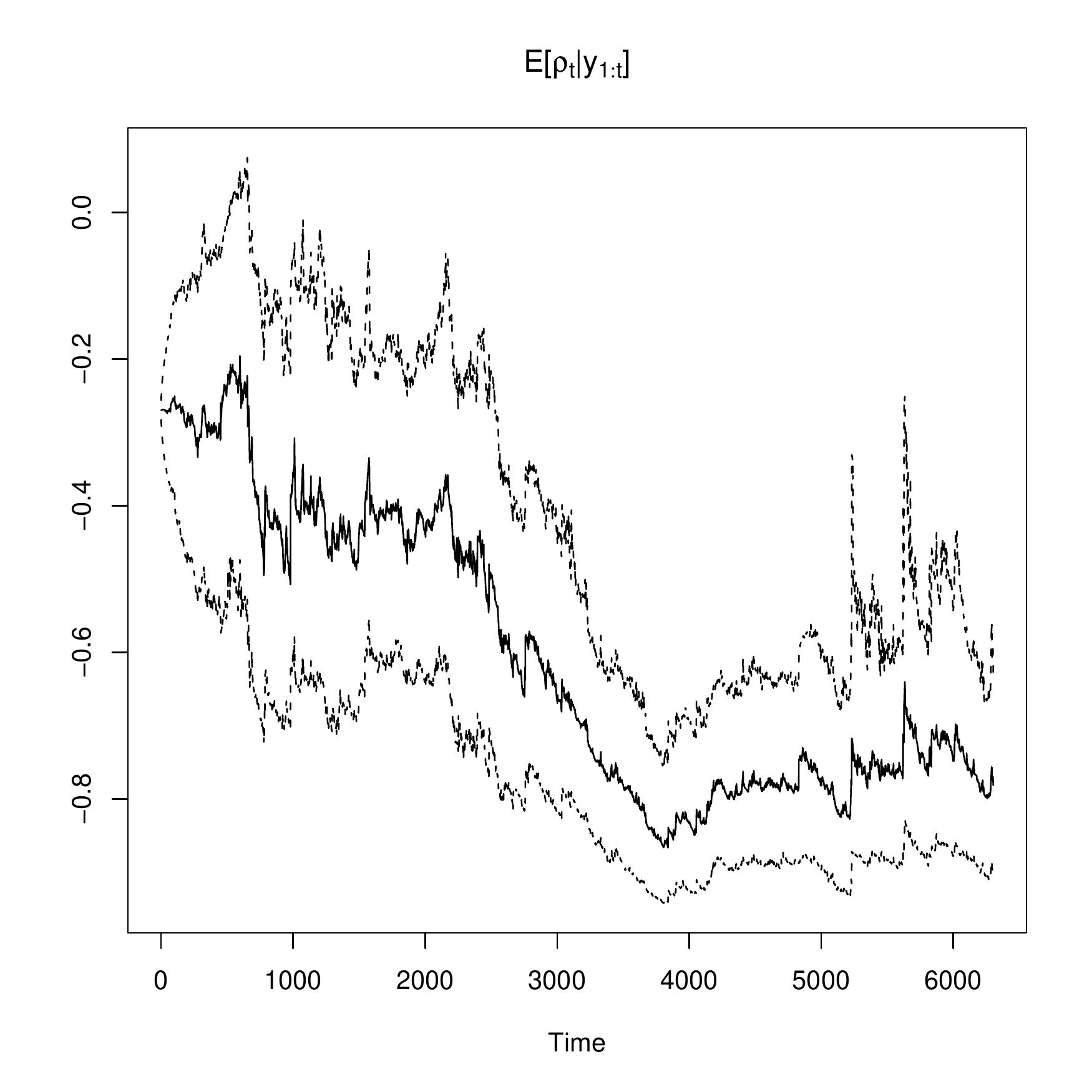}
%
\end{minipage}
{\bf \caption{\textnormal{\emph{Left panel}: Data and filtering means of some unobserved variables for the random-walk leverage model (top to bottom): S\&P500 daily returns, leverage correlation $\rho_{t}$, volatility factor $h_{t}$ and return shock $\epsilon_{t}$. 
\emph{Right panel}: mean (solid line), first quartile and third quartile (dashed lines) of the filtering distributions of the unobserved leverage correlation $\rho_{t}$ (conditional on the contemporaneously available data $y_{1},\ldots, y_{t}$). 
All results were obtained using a particle filter with 5,000 particles.}} }
\end{figure}
%
%
\section{Discussion}
\label{sec:discussion}
The new \kwp{evidence} reported here supporting random leverage lies in the \kwy{likelihood difference} and in the \kwo{interval estimates} for the unobserved leverage $\rho_{t}$ and both strengthens the current, scant empirical support for general random leverage and \kwb{complements existing theoretical work} on idiosyncratic leverage. 
The reported \kwy{likelihood} of the random-walk leverage model is higher than that of the fixed-leverage one. 
A formal model comparison may be carried using a model selection criterion, like Akaike's information criterion.  
Any such criteria that penalizes higher likelihoods according to some increasing function of the number of estimated parameters will favor the random-walk leverage model, since the number of parameters in both models is four. 
However, these models are not nested and cannot be compared with a likelihood-ratio test. 
Still, for such test to fail to deem significant at a 5\% level the improvement of 6.78 log-likelihood units that Table~1 reports, it would take a model with at least seven more parameters than the fixed-leverage model. 
This number of additional allowed parameters gives an informal measure of how higher the ability to capture the dynamics in the data of the random-walk leverage model is. 
Other informal support in favor of the random-walk leverage model is given by the interval estimates for $\rho_{t}$ reported in Figure~1. 
The upper and lower bounds of these intervals evolve more or less monotonically from approximately $(-0.55, -0.1)$ 
down to approximately $(-0.6, -0.9)$. 
These intervals intersect during a large fraction of the period considered with the 95\% confidence interval reported for the fixed-leverage model $\rho$ of $(-0.7399, -0.5759)$. 
Nevertheless, this intersection does not occur (or is minimal) in some sub-periods, supporting time-variation even if uncertainty about $\rho_{t}$ is taken into account. 
Ignoring this uncertainty and focusing on the filtering mean of $\rho_{t}$ also provides information about leverage stationarity, which has been assumed in previous theoretical work regarding idiosyncratic random leverage. 

The leverage stationarity assumption of \cite{veraart2012} suggests that the filtered $\rho_{t}$ should oscillate around a mean leverage value. 
Such oscillations would take longer times as parameters approached non-stationarity regions and hence (for a fixed sample size) $\rho_{t}$ would return fewer times to its mean within the sample. 
Such mean-reverting behavior seems to be supported by the filtering means in Figure~1 but only for some sub-periods of our sample (e.g., observations 500 to 2000, or 4500 to 6000) and not for the entire sample. 
This suggests the following two conjectures (left to be explored in future research): first, a stationary model might be more appropriate for certain sub-periods than for the entire period considered here; and second, if a stationary model were used (instead of a random walk) for the entire sample, parameters would likely be estimated to be near non-stationarity regions (in fact, this was observed at preliminary stages of our work when fitting a stationary AR(1) leverage model to the entire sample). 
Such closeness to non-stationarity regions would raise the question of testing for non-stationarity. 
The need for such tests has given rise to an abundant literature that extends the seminal work on unit root testing of \cite{dickey1979}. 
Unit roots have also been considered in the state-space framework, having produced so far a careful study of diffuse priors for Gaussian exact filtering and smooting \citep[see for example][]{durbin2001} and of unit-root testing in Gaussian, linear state-space models \citep{chang2009}, but no results seem to have been developed so far for the non-linear, non-Gaussian framework needed in this paper. 

%
%
¥
\section*{Acknowledgements}
This work was supported by Spanish Government Project ECO2012-32401 and Spanish Program \emph{Juan de la Cierva} (JCI-2010-06898). 
%
\bibliographystyle{elsarticle-harv}
%
\bibliography{references}
%
%
%
%
%

\newpage

\appendix
\renewcommand{\thesection}{Appendix \Alph{section}}
\renewcommand{\theequation}{\Alph{section}.\arabic{equation}}
\setcounter{equation}{0}
\setcounter{section}{0}
%
%
¥
\section{}
\label{app:diagnostics}

\begin{figure}[ht]
\centering
\begin{minipage}[b]{0.33\linewidth}
  \includegraphics[width=1\linewidth]{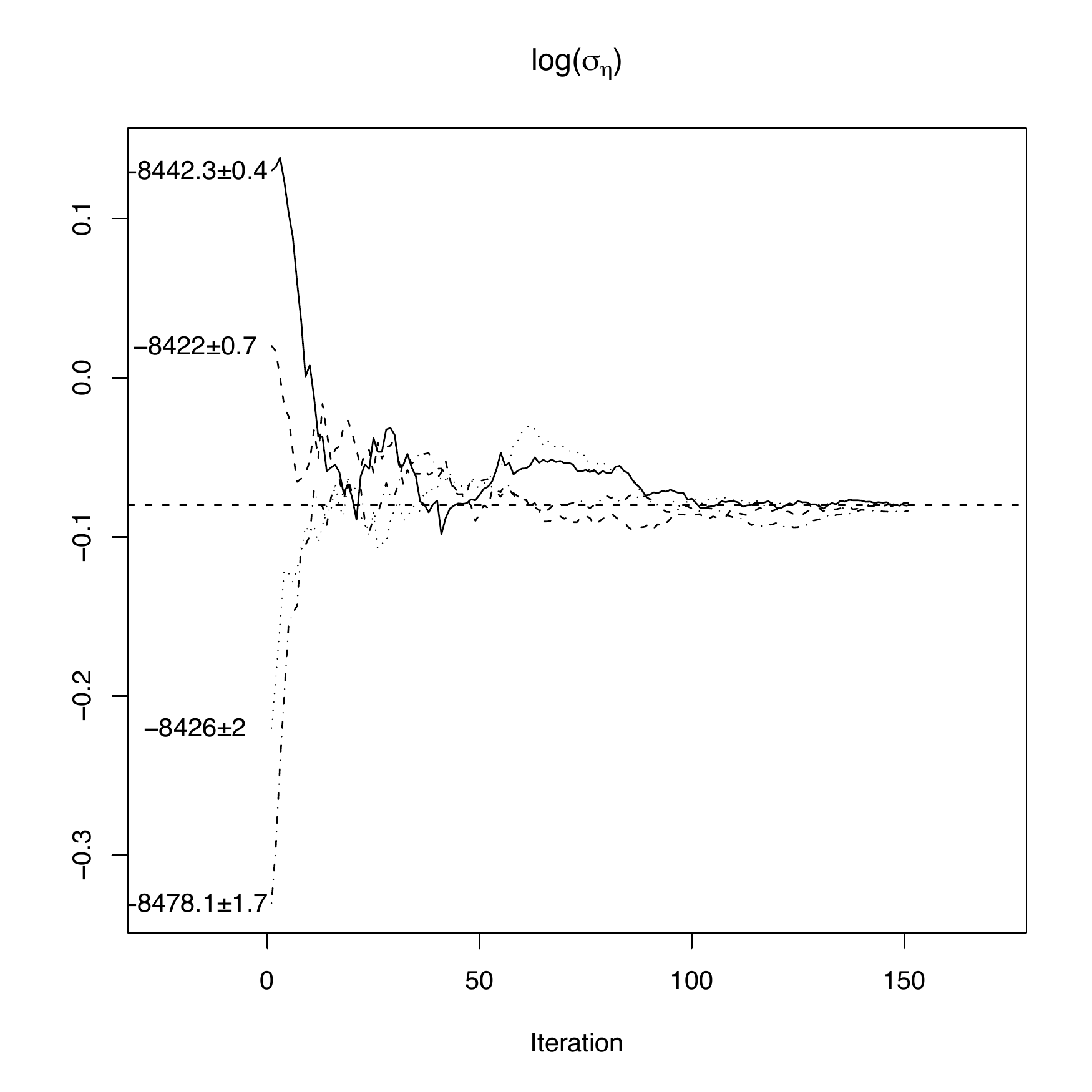}
\end{minipage}
\hspace{-0.8cm}
\begin{minipage}[b]{0.33\linewidth}
  \includegraphics[width=1\linewidth]{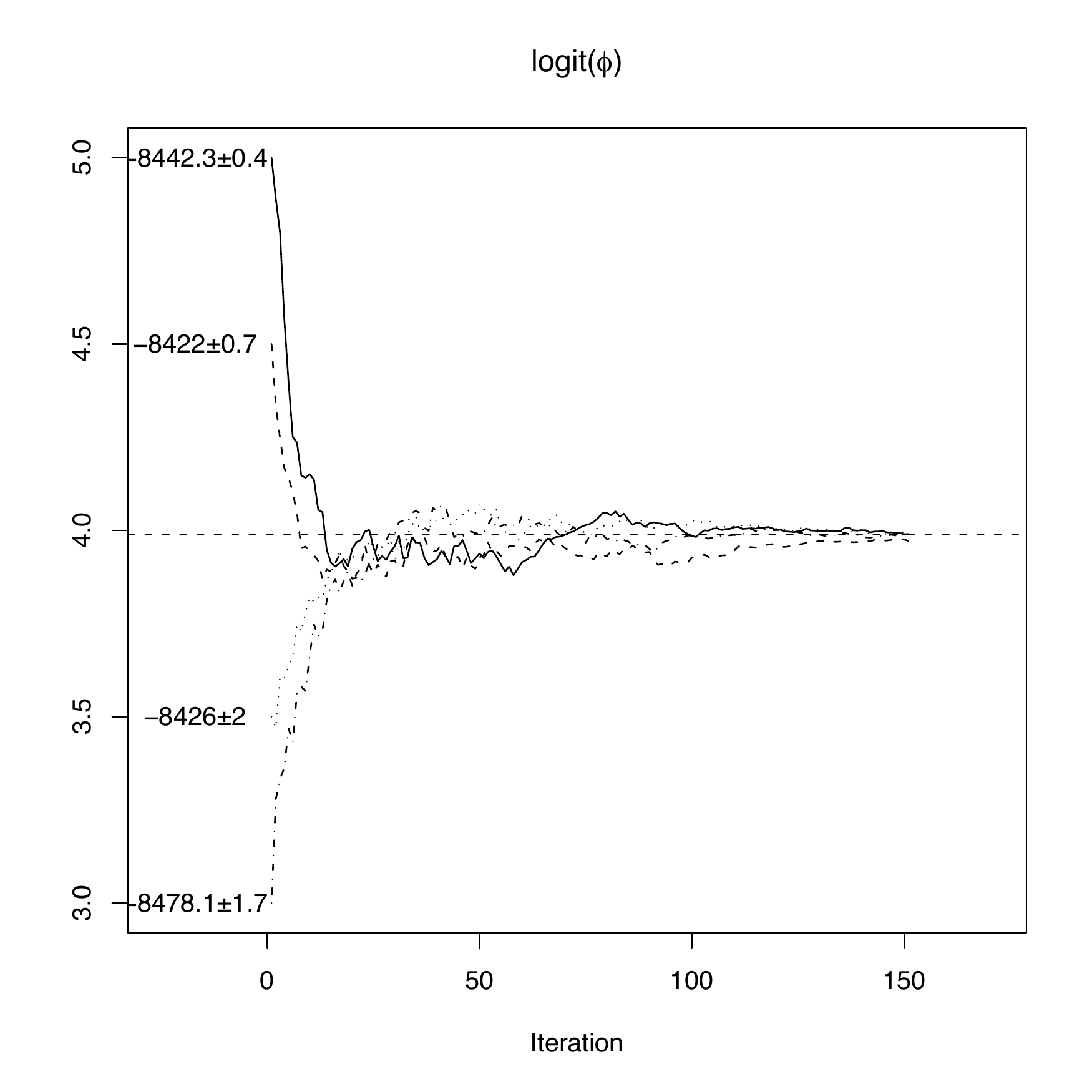}
\end{minipage}
\hspace{-0.8cm}
\begin{minipage}[b]{0.33\linewidth}
  \includegraphics[width=1\linewidth]{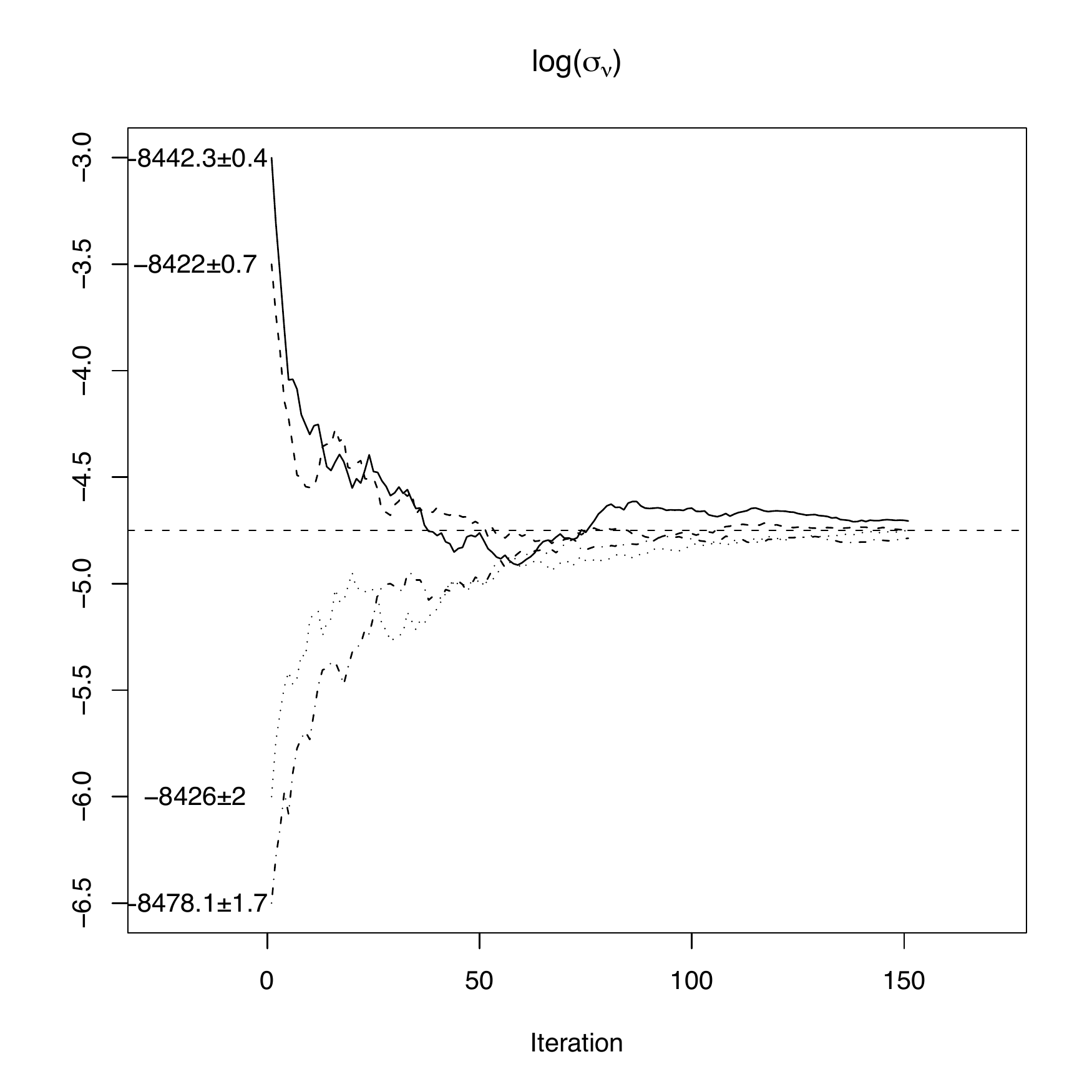}
\end{minipage}
\begin{minipage}[b]{0.33\linewidth}
  \includegraphics[width=1\linewidth]{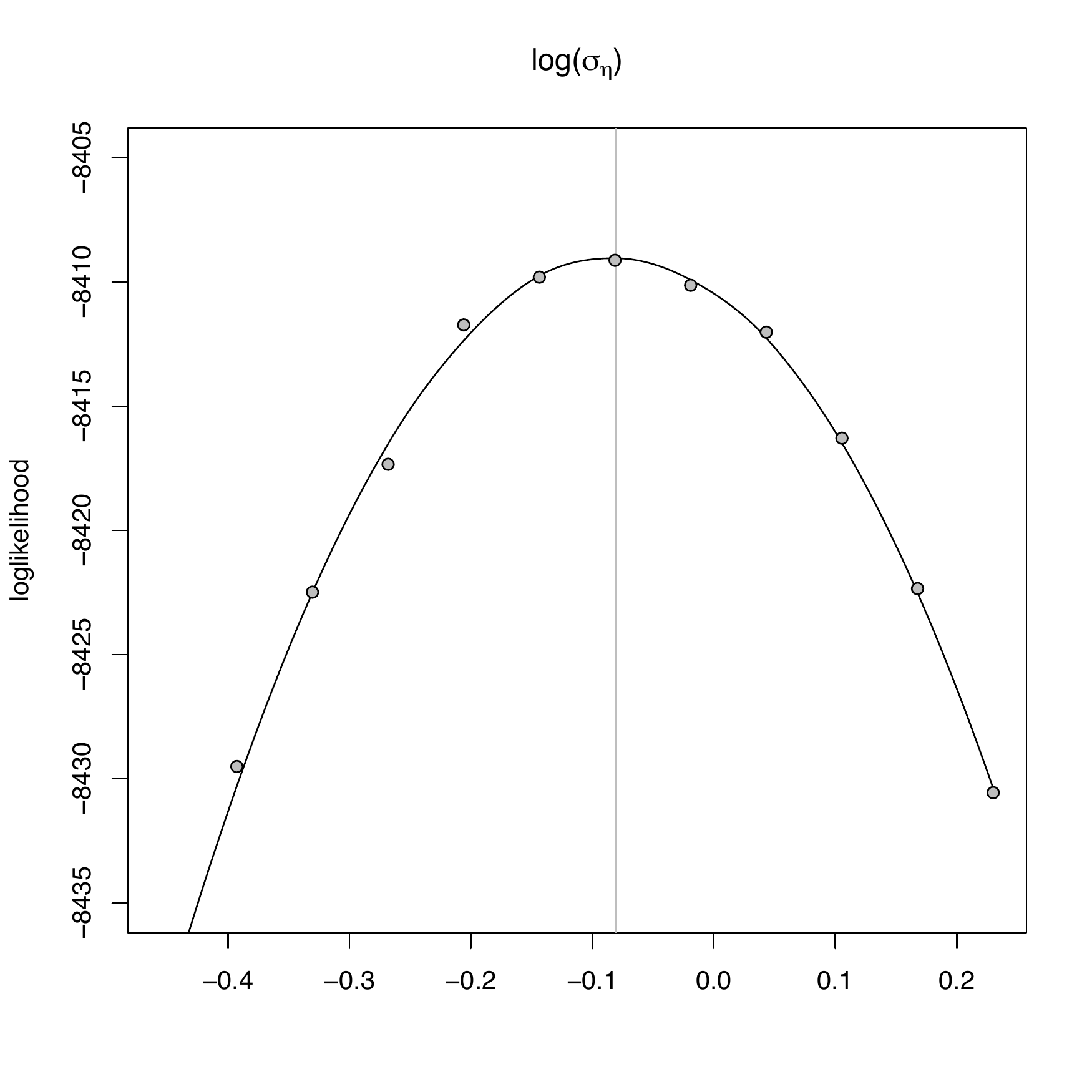}
\end{minipage}
\hspace{-0.4cm}
\begin{minipage}[b]{0.33\linewidth}
  \includegraphics[width=1\linewidth]{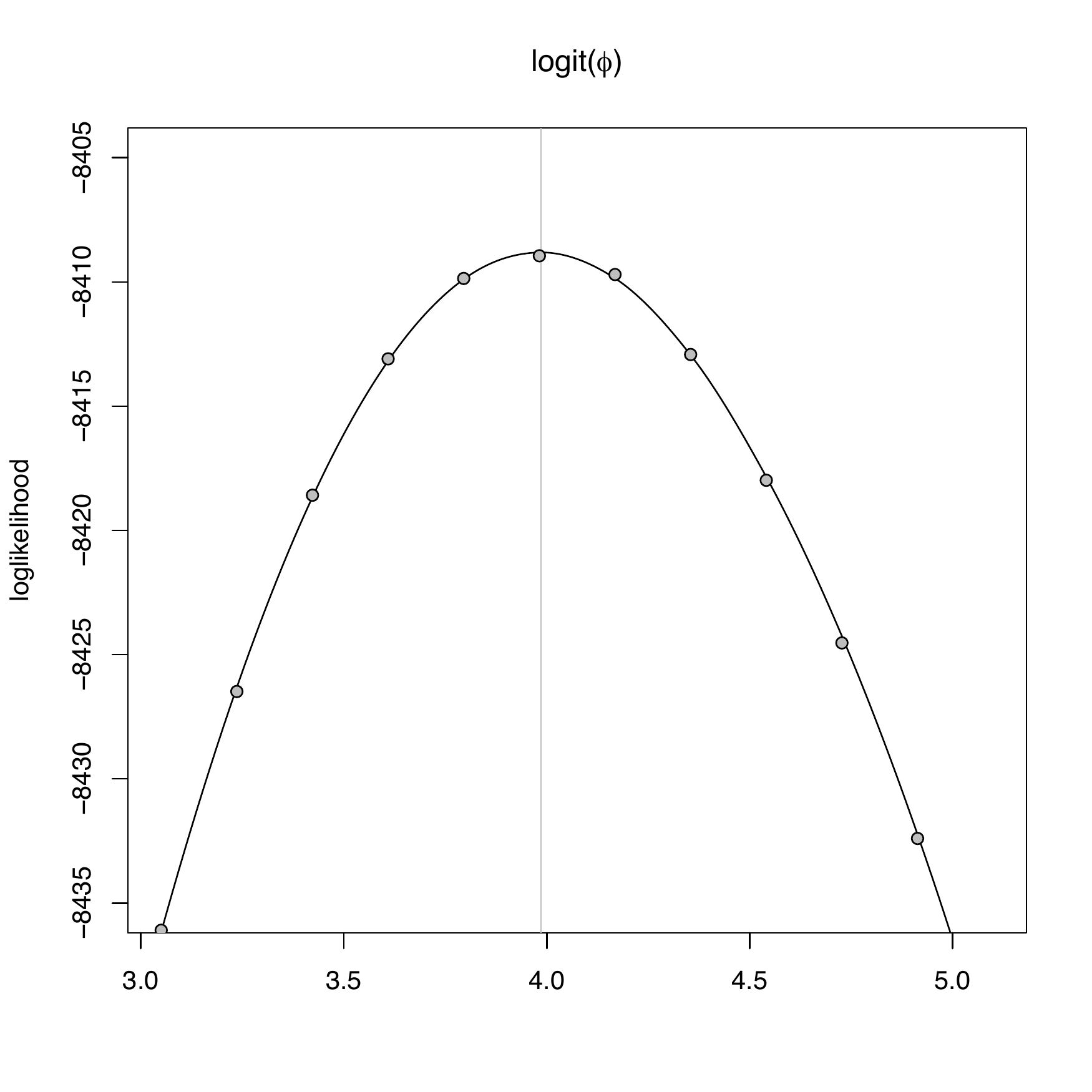}
\end{minipage}
\hspace{-0.4cm}
\begin{minipage}[b]{0.33\linewidth}
  \includegraphics[width=1\linewidth]{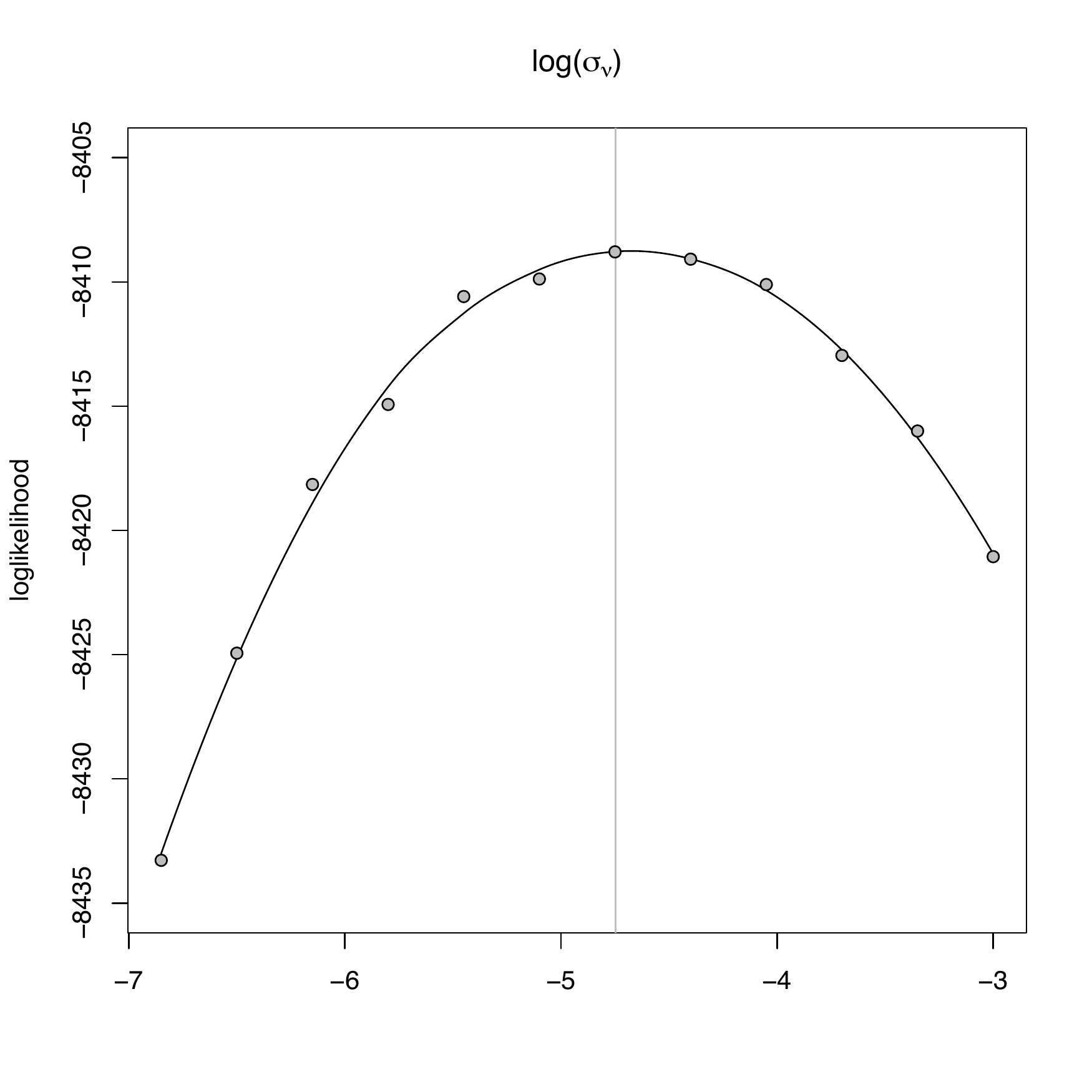}
\end{minipage}

{\bf \caption{ \textnormal{ 
Sample of diagnostic plots for iterated filtering corresponding to transformations (to $\R$) of parameters of random-walk leverage model: $\log(\sigma_{\eta})$ (left column), $\text{logit}(\phi)$ (center column), and $\log(\sigma_{\nu})$ (right column).  
\emph{Top panels}: trajectories of the algorithm updates showing $\hat{\theta}_{1}, \hat{\theta}_{2}, \ldots, \hat{\theta}_{150}$ converge to the maximum likelihood estimate (horizontal line) starting from four different starting points $\hat{\theta}_{0}$. 
On the left margin, likelihood estimates at the starting point $\hat{\theta}_{0}$ and Monte Carlo interval half width, obtained by averaging four replicates with 10,000 particles each. 
\emph{Bottom panel}: sliced likelihoods for the corresponding parameters, in which the likelihood surface is explored along one of the parameters, keeping the other parameters fixed at the point which iterated filtering converges to. 
Each filled circle shows the likelihood estimate obtained with 70,000 particles. 
The curve results from smoothing the likelihood evaluations with local quadratic regression. 
}}}
\end{figure}

\end{document}
%

%